\documentclass[aps,groupedaddress,amsmath,preprintnumbers,twocolumn]{revtex4}
\input{amssym}
\usepackage{hyperref}

\def\be {\begin{equation}}
                                                                                
\def\ee  {\end{equation}}
                                                                                
\def\bea {\begin{eqnarray}}
                                                                                
\def\eea {\end{eqnarray}}
                                                                                
\def\nn {\nonumber}
\begin{document}
\preprint{ }
\title{How red is a quantum black hole?}
\author{Viqar Husain and Oliver Winkler}
\email[]{husain@math.unb.ca, oliver@math.unb.ca}
\affiliation{Department of Mathematics and Statistics,
University of New Brunswick, Fredericton, NB, Canada E3B 5A3}
\date{\today}

\begin{abstract}

Radiating black holes pose a number of puzzles for semiclassical 
and quantum gravity. These include the transplanckian  problem -- the 
nearly infinite energies of Hawking particles created near the horizon, 
and the final state of evaporation. A definitive resolution 
of these questions likely requires robust inputs from quantum gravity. 
We argue that one such input is a quantum bound on curvature. We show 
how this leads to an upper limit on the redshift of a Hawking emitted 
particle, to a maximum temperature for a black hole, and to the prediction 
of a Planck scale remnant.
 
\end{abstract}
                                                                                
\maketitle
\vfill\eject
                                                                               
Hawking radiation \cite{hawk, unruh} has placed the analogy of black hole 
physics with thermodynamics on a firm basis \cite{bd,rev}, while at the same 
time posing a challenging set of questions for quantum gravity. These include a
deeper understanding of the origin of emitted particles, black hole entropy, and  
the end point of evaporation. 

The first question is significant already at the semiclassical level --  
a Hawking mode detected far from a black hole must have redshifted 
from a far higher frequency near the horizon. The closer the origin of 
the mode to the horizon, the higher the energy with which it must have 
been created. This suggests the availability of an infinite transplanckian 
reservoir of states. It also presents a problem: if emitted 
particles originate with transplankian frequencies, their backreaction 
effects are not negligible, and the semiclassical approximation breaks 
down. No final solution of this problem is acknowledged, although it has been 
shown that Hawking radiation persists if Lorentz invariance violating 
dispersion relations are introduced for which high frequency modes do not 
propagate \cite{bill,ted}.  
  
The black hole entropy question has been addressed in various approaches to quantum gravity, 
but all the results so far have clear limitations. The string theory solution \cite{stringbh} 
applies to highly special extremal or "near" extremal black holes. Although it incorporates 
"dynamics" in the sense that the quantum states counted belong to a supersymmetry 
multiplet (which necessarily incorporates a Hamiltonian), it provides no understanding 
of the entropy of the simplest Schwarzschild black hole. Other solutions, 
such as that in loop quantum gravity \cite{isoh} and boundary conformal field 
theory \cite{steve}, identify degrees of freedom on a horizon, which is taken 
to be a fixed classical entity. This obviously has its limitations, the main 
one being that any type of classical horizon boundary condition freezes what 
would otherwise be interesting quantum fluctuations. Indeed, quantum gravity would 
be rather  dull with a classically frozen conformal structure. 

Not surprisingly, various models have been utilised over the years to study the 
third question as well \cite{gidd}. There are three general scenarios: evaporation leaves 
behind a flat or other non-black hole spacetime, a Planck scale remnant, or a tunnel to a 
new universe. The information lost into the black hole during its lifetime in these 
is, respectively, recovered, stored in the remnant, or has escaped to the new 
universe. 

What appears to be missing in the details of the various solutions to these 
questions is an identification of one or more central and robust features that 
are expected to emerge from any quantum gravity theory. We are not refering 
here to basic intuitive ideas, such as a fundamental length scale, foamy 
spacetime, and so on, but rather to concrete mathematical results that 
one can argue are model independent. We identify here one such robust feature, namely  
{\it a curvature operator with bounded spectrum}. We show how this property alone 
can address the transplanckian mode and final state problems.  

The setting for our observations is the spherically symmetric gravity-scalar field 
system. This is the "standard model" to study, since {\it all} of the 
semi-classical results of black hole physics originate here. Indeed one can argue 
that a complete quantization of this system is a fundamental problem, whose solution 
would result in a complete understanding of quantum black hole physics.   
 
Recently we have initiated work \cite{hw1,hw2} on the quantization of this 
system. The dynamical variables are a metric function $R(r,t)$ whose square gives 
the areas of spheres, and a massless scalar field $\Phi(r,t)$. The basis states 
of the quantization are such that these dynamical variables are diagonal:
\bea 
   \hat{R}|r,\phi\rangle &=& \sqrt{2}l_P r\ |r,\phi\rangle, \nn\\
   \hat{\Phi}|r,\phi\rangle &=& \phi\ |r,\phi\rangle, 
\eea
where $r,\phi$ are dimensionless numbers and $l_P$ is the Planck length.  
In addition there is an inverse $R$ operator whose action on every basis state 
is well defined, given by \cite{hw1}
\bea 
\widehat{\frac{1}{R}}\ |r,\phi\rangle 
= \frac{\sqrt{2}}{l_P}\ \left| \sqrt{|r|}-\sqrt{|r-1|}\right ||r,\phi\rangle. 
\label{1/R}
\eea 
Important features of this result are that the spectrum is bounded above, 
with largest eigenvalue $\sqrt{2}/l_P$, and that for large values of $r$, the 
eigenvalue behaves like $1/r$, as it should for classical behaviour.  
Thus, this spectrum exhibits quantum corrections for short distances, and 
recovers classical behaviour for large distances.  

An important physical property of (\ref{1/R}) follows from the observation that 
classically all curvature variables in spherical symmetry are proportional to 
positive powers of $1/r$. This means that the corresponding curvature operators, 
which are constructed from products of this operator are all bounded above. 
This is the robust result whose consequences we now explore.  

Let us consider a black hole of mass $M$ and horizon radius $R$. This radius is related 
to the mass by the relation $R=2M$. The essence of the problem of infinite redshifts, 
known as the transplanckian problem, is that a Hawking mode of frequency $\omega_o$, 
observed far from the horizon of a black hole at location $R_o$, originated near the 
horizon at radial location $R_c$ with frequency  
\be 
   \omega_c = Z \omega_o = \sqrt{\frac{1-2M/R_o}{1-2M/R_c}} \ \omega_o.
\ee
For $R_c$ arbitrarily close to the value $2M$, the created particles have 
energies that are larger than the Planck energy $M_P=1.2\times 10^{19}$GeV. 

The question for quantum gravity is whether there is a quantum analog of the 
classical redshift variable $Z$, constructed from the dynamical variables and their 
momenta, whose spectrum is bounded above. This would provide a quantum gravity basis 
for an upper limit on the frequency of created particles, and perhaps also for the 
ad-hoc modified dispersion relations discussed in earlier work \cite{bill,ted}. 

It turns out that one can construct such a quantum variable. Let us consider for 
simplicity the case $R_o \rightarrow \infty$ for a Hawking particle detected at 
infinity. Then the redshift factor is 
\be 
 Z_\infty = \sqrt{\frac{R_c}{{R_c-2M}}},
\ee 
where $M=m M_{Planck}$ is the (constant) mass associated with a given black hole. 
Motivated by (\ref{1/R}), which is ultimately connected to a classical phase space 
expression, we can define an operator in the fixed mass sector of the theory 
corresponding to the classical function $Z_{\infty}$. This operator has the spectrum   
\be 
 \sqrt{2|r|} \ 
 \left| |r-2m|^{1/2}-|r-2m-1|^{1/2}\right|
\ee
This expression has  desirable  correspondences: for $r=2m+1$ the spectrum has a maximum, 
whereas for $r>>2m$, the eigenvalue reduces to the classical expression. The interesting 
result for the transplanckian problem is that the redshift operator has maximum 
eigenvalue  
\be 
 z_{\rm max} = \sqrt{2(2m+1)}. 
\ee
This quantum gravity result has the physical consequence that for Planck scale black 
holes, there is a much smaller frequency gap between created and observed particles than 
for larger black holes. For example, for a 10 solar mass black hole this formula gives  
$z_{max} = 6.3\times 10^{19}$, whereas for one that is a few Planck masses the number is 
of order $10$. 

Our second observation motivated by (\ref{1/R}) relates to the end point of black 
hole evaporation. It is well known that the semi-classical analogy of black hole physics 
with thermodynamics gives the temperature relation
\be 
  T = \frac{1}{8\pi M} = \frac{1}{4\pi R_H}.
\ee 
In contrast to normal thermodynamics, this relation indicates that temperature is a phase 
space variable in black hole mechanics, and can therefore be represented as an operator 
upon quantization. Furthermore, this relation must arise from a full quantisation for 
semi-classical states describing a fixed mass sector. Since such states are linear combinations
of basis states, the upper bound on the spectrum (\ref{1/R}) is unaffected.    

The interesting observation now is that this upper bound leads to a maximum temperature. We are 
thus led to a quantum gravity modification of the Hawking temperature formula given by  
\be 
 T = \frac{\sqrt{2}}{4\pi l_P}\ \left(\sqrt{r_H}-\sqrt{|r_H-1|}\right).
\ee

The consequence of this result for the end point of evaporation is now clear. As a 
black hole shrinks, the maximum temperature $T_{\rm max} = 1.6\times 10^{31}$ Kelvin 
is achieved at $r_H=1$ (in Planck units), and zero temperature at $r_H=1/2$. Furthermore, 
the specific heat determined by this formula turns from negative to positive at $r_H=1$, 
and becomes zero at $r_H=1/2$. These features suggest a Planck scale remnant of size 
$l_P/\sqrt{2}$ as the last stage of evaporation.  

In summary, we have observed that a quantization of gravity in which curvature operators 
have a bounded spectrum leads to some dramatic conclusions for  important open questions 
of black hole physics: they predict a mass dependent upper bound on the gravitational 
redshift, a maximum temperature for black holes, and a cold remnant as the end point 
of evaporation.  

\medskip

\noindent {\it Acknowledgement} This work was supported in part by the Natural Science
and Engineering Research Council of Canada.

\end{document}